\documentclass[a4paper, 10pt, unnumberedsections, twoside]{LTJournalArticle}

\usepackage[farskip=0pt]{subfig}
\usepackage{lipsum}
\usepackage[acronym]{glossaries}
\usepackage[per-mode=symbol,per-symbol =/]{siunitx}
\usepackage[capitalise]{cleveref}

\DeclareSIUnit{\x}{\!\ensuremath{\times}}
\DeclareSIUnit\bit{b}
\DeclareSIUnit\gateeq{GE}
\runninghead{}
\footertext{\textit{RISC-V Summit Europe, Munich, 24-28 June 2024}}
\title{SentryCore: A RISC-V Co-Processor System for Safe, Real-Time Control Applications}

\ifx\blind\undefined
    \author{%
        Michael Rogenmoser,\textsuperscript{1}\thanks{Corresponding author: \href{mailto:michaero@iis.ee.ethz.ch}{\tt michaero@iis.ee.ethz.ch}\\This work was supported in part through the TRISTAN (101095947) project that received funding from the HORIZON CHIPS-JU programme} \
        Alessandro Ottaviano,\textsuperscript{1} \
        Thomas Benz,\textsuperscript{1} \\\
        Robert Balas,\textsuperscript{1} \
        Matteo Perotti,\textsuperscript{1} \
        Angelo Garofalo,\textsuperscript{1,2} \
        Luca Benini\textsuperscript{1,2}
    }
\else
    \author{\textit{Authors omitted for blind review}}
\fi

\newacronym{fpga}{FPGA}{field programmable gate array}
\newacronym{asic}{ASIC}{application-specific integrated circuit}
\newacronym{fub}{FUB}{functional unit block}
\newacronym{vv}{V\&V}{validation and verification}
\newacronym{gpp}{GPP}{general purpose processor}
\newacronym{tid}{tID}{transaction ID}
\newacronym{ai}{AI}{artificial intelligence}

\newacronym{hpc}{HPC}{high performance computing}
\newacronym{ml}{ML}{machine learning}
\newacronym{isa}{ISA}{instruction set architecture}
\newacronym{fp}{FP}{floating-point}
\newacronym{dl}{DL}{deep learning}
\newacronym{la}{LA}{linear algebra}
\newacronym{ip}{IP}{intellectual property}
\newacronym{mpsoc}{MPSoC}{multi-processor system-on-chip}
\newacronym[firstplural=networks-on-chip (NoCs)]{noc}{NoC}{network-on-chip}
\newacronym{swapc}{SWaP-C}{space, weight, power, and cost}
\newacronym{mcp}{MCP}{multi-core processor}
\newacronym{rr}{RR}{round-robin}
\newacronym{mcu}{MCU}{microcontroller unit}

\newacronym{mac}{MAC}{multiply-accumulate}
\newacronym{fem}{FEM}{finite element analysis}
\newacronym{simd}{SIMD}{single-instruction, multiple-data}
\newacronym{rtl}{RTL}{register transfer level}
\newacronym{dlt}{DLT}{data layout transform}

\newacronym{fifo}{FIFO}{first in, first out}
\newacronym{fu}{FU}{functional unit}
\newacronym{alu}{ALU}{arithmetic logic unit}
\newacronym{ssr}{SSR}{stream semantic register}
\newacronym{issr}{ISSR}{indirection stream semantic register}
\newacronym{tcdm}{TCDM}{tightly-coupled data memory}
\newacronym{dma}{DMA}{direct memory access}
\newacronym{sm}{SM}{streaming multiprocessor}
\newacronym{vlsu}{VLSU}{vector load-store unit}
\newacronym{dsa}{DSA}{domain-specific accelerator}
\newacronym{ha}{HA}{hardware accelerator}
\newacronym{fsm}{FSM}{finite state machine}
\newacronym{llc}{LLC}{last-level cache}
\newacronym{d2d}{D2D}{die-to-die}
\newacronym{dram}{DRAM}{dynamic random access memory}
\newacronym{spm}{SPM}{scratchpad memory}
\newacronym{rf}{RF}{register file}
\newacronym{mmu}{MMU}{memory management unit}

\newacronym{os}{OS}{operating system}
\newacronym{gpos}{GPOS}{general-purpose operating system}

\newacronym{spvv}{SpVV}{sparse vector-vector multiplication}
\newacronym{spmv}{SpMV}{sparse vector-matrix multiplication}
\newacronym{spmm}{SpMM}{sparse matrix-matrix multiplication}
\newacronym{csrmv}{CsrMV}{CSR matrix-vector multiplication}
\newacronym{csrmm}{CsrMM}{CSR matrix-matrix multiplication}

\newacronym{csf}{CSF}{compressed sparse fiber}
\newacronym{csr}{CSR}{compressed sparse rows}
\newacronym{csc}{CSC}{compressed sparse columns}
\newacronym{bcsr}{BCSR}{blocked compressed sparse rows}

\newacronym{axi4}{AXI4}{Advanced eXtensible Interface 4}
\newacronym{amba}{AMBA}{Advanced Microcontroller Bus Architecture}
\newacronym{sram}{SRAM}{static random-access memory}

\newacronym{wcet}{WCET}{worst-case execution time}
\newacronym{rtunit}{REALM unit}{real-time regulation and traffic monitoring unit}
\newacronym{mtunit}{M\&R unit}{monitoring and regulation unit}
\newacronym{cps}{CPS}{cyber-physical system}
\newacronym{crtes}{CRTES}{critical real-time embedded system}
\newacronym{heicps}{He-iCPS}{heterogeneous integrated cyber-physical system}
\newacronym{ecu}{ECU}{electronic control unit}
\newacronym{mcs}{MCS}{mixed-criticality system}
\newacronym{ima}{IMA}{integrated modular avionics}
\newacronym{axirealm}{AXI-REALM}{AXI real-time regulation and traffic monitoring}
\newacronym{mpam}{MPAM}{memory system resource partitioning and monitoring}
\newacronym{dos}{DoS}{denial of service}
\newacronym{hwrot}{HWRoT}{hardware root of trust}
\newacronym{pcs}{PCS}{power controller subsystem}
\newacronym{sil}{SIL}{safety integrity level}
\newacronym{obi}{OBI}{open bus interface}
\newacronym{pmca}{PMCA}{programmable many-core accelerator}
\newacronym{ap}{AP}{application-class processor}
\newacronym{clic}{CLIC}{core-local interrupt controller}
\newacronym[plural={RTOSes}, \glsshortpluralkey={RTOSes}]{rtos}{RTOS}{real-time operating system}

\newacronym{tcls}{TCLS}{triple-core lockstep}
\newacronym{ecc}{ECC}{error-correcting code}

\ifx\showacrcols\undefined
    
\else
    
\fi

\ifx\blind\undefined
    \date{
        \vspace{-0.32em}
        \footnotesize\textsuperscript{\textbf{1}}Integrated Systems Laboratory, ETH Zurich \\
        \footnotesize\textsuperscript{\textbf{2}}Department of Electrical, Electronic, and
        Information Engineering, University of Bologna
        \vspace{-0.32em}
    }
\else
    \date{}
    \vspace{1.1cm}
\fi

\renewcommand{\maketitlehookd}{%
\begin{abstract}
In the last decade, we have witnessed exponential growth in the complexity of control systems for safety-critical applications (automotive, robots, industrial automation) and their transition to heterogeneous \glspl{mcs}.
The growth of the RISC-V ecosystem is creating a major opportunity to develop open-source, vendor-neutral reference platforms for safety-critical computing.
We present SentryCore, a reliable, real-time, self-contained, open-source mega-IP for advanced control functions that can be seamlessly integrated into Systems-on-Chip, e.g., for automotive applications, through industry-standard \gls{axi4}. 
SentryCore features three embedded RISC-V processor cores in lockstep with \gls{ecc} protected data memory for reliable execution of any safety-critical application. 
Context switching is accelerated to under 110 clock cycles via a RISC-V \gls{clic} and dedicated hardware extensions, while a timer-based \gls{dma} engine streamlines sensor data readout during periodic control loops.
SentryCore was implemented in Intel's 16nm process node and tested with FreeRTOS, ThreadX, and RTIC software support.
\end{abstract}
\vspace{-0.3cm}
}

\begin{document}

\maketitle
\glsresetall

\section{Introduction}
\vspace{-0.1cm}


With the increase in complexity of application domains such as automotive, robotics, avionics, and space, modern edge computing systems demand the integration of heterogeneous functionalities with different criticality levels. 
These \glspl{mcs} require competitive peak performance for non-critical workloads while guaranteeing safety, reliability, and real-time properties for safety- and time-critical tasks.


\emph{Isolation} is the state-of-the-art technique to achieve freedom from interference on heterogenous \glspl{mcs}~\cite{MCS_REVIEW_2}. 
In particular, \emph{physical} isolation relies on federated hardware for each software component: non-critical, computationally-intensive workloads --- e.g., image processing in advanced driver-assistance systems --- are executed on application-class processors running general-purpose OSes, enhanced with targeted accelerators; safety- and time-critical control tasks --- e.g., brake control in cars or attitude control in satellites --- are executed by dependable co-processor architectures with \gls{rtos} support.



Despite their compact size, akin to commodity microcontroller units, the design of these co-processors is challenging since they require both dependability and predictability features.
Latency must be minimized to ensure the execution time of critical applications is tightly bounded. 
For the reliability of the co-processor, no exotic technology choices, such as radiation-hardened nodes, process design kits, or standard cells, can be made for the critical system, as these would hamper the performance of the full \gls{mcs}. Therefore, architectural mechanisms such as triple modular redundancy and \glspl{ecc} are required to protect against soft errors.
%
%
%
%
Closed-source solutions are available~\cite{st_stellar_2024, renesaes_rcar_2024}; however, given the momentum of RISC-V, a vendor-neutral design based on the open \gls{isa} is not yet available and could find significant market traction.
To fill this gap, we present SentryCore, a dependable 32-bit RISC-V-based mega-IP for safety-critical, real-time subsystems of MCSs. In particular:

\vspace{-0.3cm}

\begin{itemize}
    \item We design a 32-bit system around RISC-V processing cores organized in \gls{tcls} architecture, with \gls{ecc}-protected memory for reliability. The system can be plugged into any large \gls{mcs} with commodity AXI4 interfaces.
    \item We enhance SentryCore with CV32RT, tailored for faster context switching, and a specialized \gls{dma} to automate sensor data acquisition in periodic control loops.
    \item We develop a programmer-friendly software stack with support for common open-source \glspl{rtos} and provide a preliminary evaluation of its interrupt capabilities.
\end{itemize}

\begin{figure}[t]
    \centering
    \includegraphics[width=.85\columnwidth]{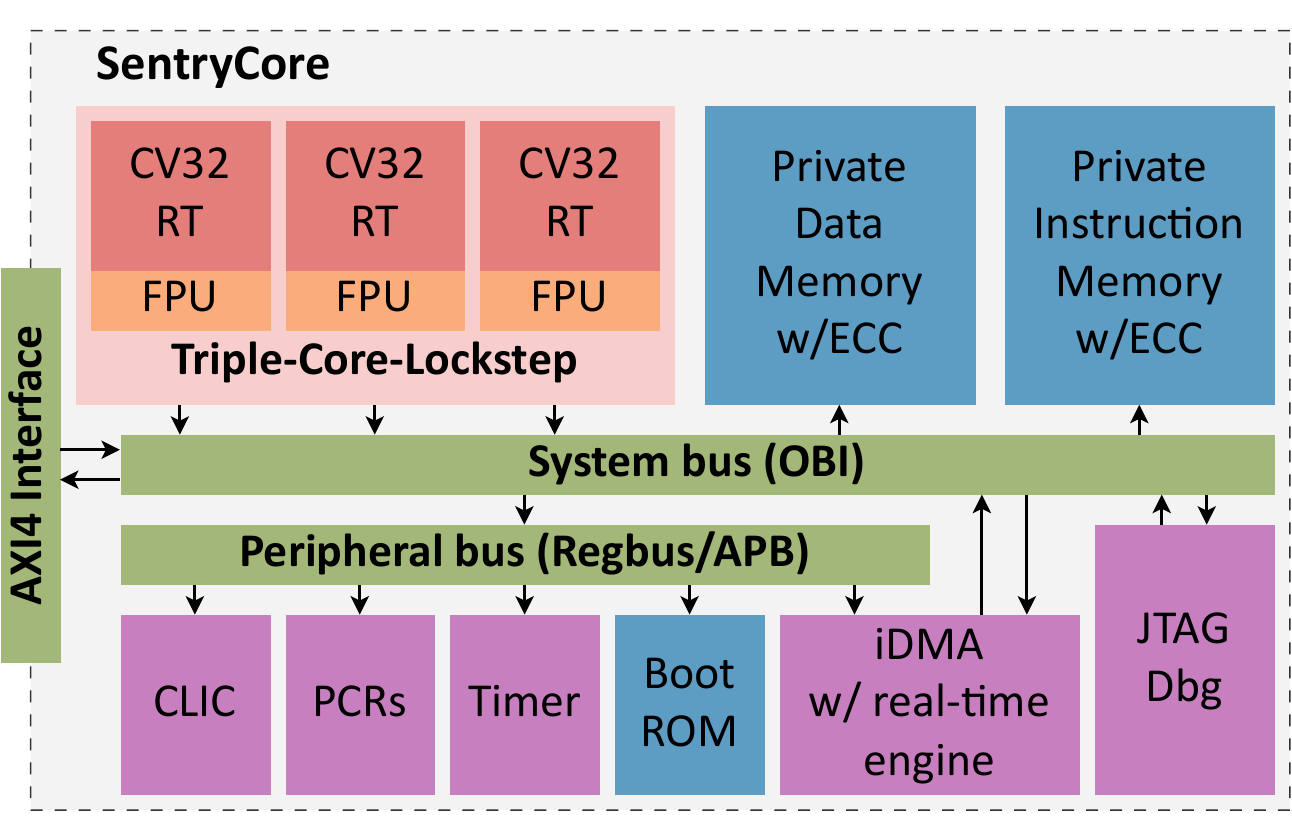}
    \caption{
        SentryCore architecture, highlighting the system’s processing cores, memory, and peripheral layout.
    }
    \label{fig:archi}
\end{figure}

\vspace{-0.3cm}
\section{Architecture}
\vspace{-0.1cm}

The open-source SentryCore co-processor mega-IP\footnote{\ifx\blind\undefined\url{https://github.com/pulp-platform/safety_island}\else URL omitted for blind review \fi} is a reliable, real-time platform built around the CV32RT~\cite{balas2023cv32rt} core, equipped with a floating-point unit critical for control tasks.
SentryCore offers significantly improved interrupt latency compared to vanilla RISC-V systems: it implements the RISC-V \gls{clic}, coupled with a custom \texttt{fastirq} extension providing register file banking and automatic context save/restore. 


For reliability, three of these cores are grouped in a \gls{tcls}~\cite{rogenmoser2023hybrid} configuration, guaranteeing fault-tolerant operations. 
Each core processes the same inputs, and their outputs undergo majority voting, ensuring system dependability even if a transient fault occurs.
To correct a mismatch within the cores' state, they start a software-based recovery mechanism, saving the correct state to the stack through the majority voter. 
The cores are then reset to clear any remaining errors, reloading the correct saved state to resume program execution within just 600 cycles of starting this resynchronization.

As shown in \cref{fig:archi}, the \gls{tcls} CV32RT cores are connected to an \gls{obi}~\cite{bink2023OBI} interconnect crossbar, with independent access to two \gls{ecc}-protected memory banks for instruction and data, with scrubbers to avoid latent errors. A further port connects to internal peripherals, namely the \gls{clic}, platform control registers, a general-purpose timer, and a local boot {ROM}. A RISC-V debug module ensures debug capabilities. 
Access into and out of SentryCore is mediated by a manager and a subordinate \gls{axi4} port.



Finally, the {iDMA} engine~\cite{10311078}, attached to the peripheral bus, features a \emph{real-time} extension to launch three-dimensional transfers on a timed schedule, thus fetching or distributing data from or to any mapped device in the \gls{mcs} in full autonomy. 
Previous work has proven this to be beneficial when periodically gathering sensor data for system management purposes~\cite{10311078}. %

\vspace{-0.3cm}
\section{Software Support and Evaluation}
\vspace{-0.1cm}

In safety- and time-critical domains, 
software components such as \glspl{rtos} must ensure deterministic performance and robust isolation and transition the system to a safe state in case of unrecoverable errors~\cite{MCS_REVIEW_2}. 
For these reasons, SentryCore supports several \glspl{rtos}, from the popular \texttt{FreeRTOS} and the open-source ASIL-D \texttt{ThreadX} to innovative frameworks such as the Rust-based real-time interrupt-driven concurrency (RTIC), rapidly gaining significant traction.

The streamlined CV32RT design allows SentryCore to achieve interrupt latencies as low as $6$ clock cycles and context switching times below $110$ clock cycles, comparing favorably to vendor-specific technologies such as the Arm Cortex-M4~\cite{balas2023cv32rt}. 

\vspace{-0.3cm}
\section{Implementation}
\vspace{-0.1cm}

\begin{figure}[t]
    \centering
    \includegraphics[width=.8\columnwidth]{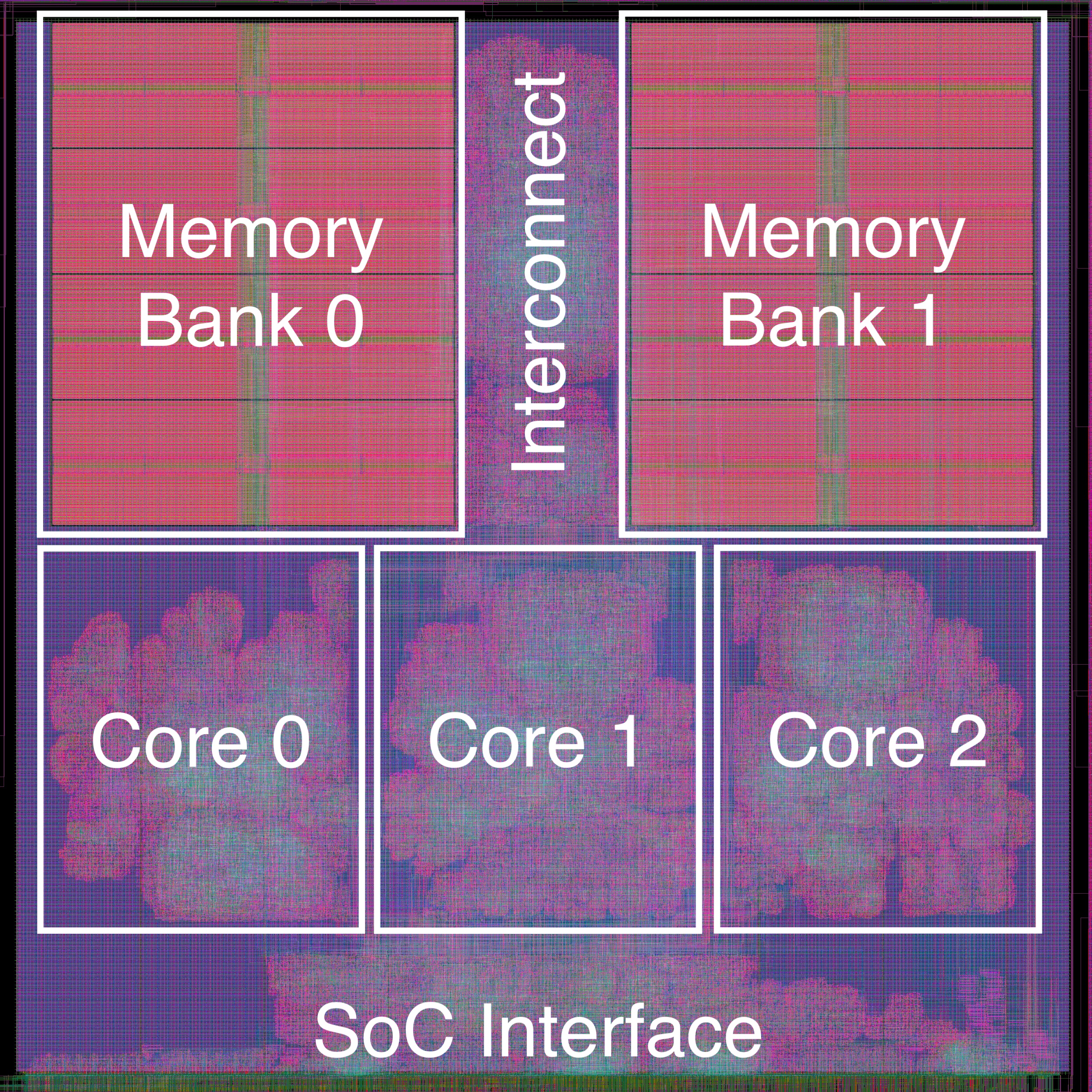}
    \caption{
        SentryCore floorplan in Intel16.
    }
    \label{fig:floorplan}
\end{figure}

SentryCore was implemented in Intel16 technology as a \SI{0.42}{\milli\meter\squared} macro. The cores are restricted to individual regions with \SI{20}{\micro\meter} margins to avoid a radiation-induced particle strike from affecting multiple cores simultaneously (\cref{fig:floorplan}).
The design is implemented at \SI{500}{\mega\hertz} and features \SI{128}{\kibi\byte} of \gls{ecc}-protected memory. A preliminary gate-level power estimation suggests a power envelope of just $50-70$~\si{\milli\watt}.





\vspace{-0.3cm}
\section{Conclusion and Future Work}
\vspace{-0.1cm}


In this paper, we presented SentryCore, an open-source, 32-bit RISC-V mega-IP designed for seamless integration in physically isolated \glspl{mcs}. 
With a streamlined interrupt architecture providing less than 110 clock cycles of context switching time, \gls{ecc} memory protection, and lockstep execution, it can bring the benefits of \gls{isa} and hardware openness to safety-critical application domains such as automotive, robotics, avionics, and space, thus enabling license-free exploration of next-generation \glspl{mcs}.

Extending reliability to the remaining components, such as the interconnect, will be handled in future work and currently relies on an external watchdog.





\vspace{-0.3cm}
{
\printbibliography}

\end{document}